%% file: paper.tex
\documentclass[11pt]{article}

\usepackage[preprint]{acl}

\usepackage{times}
\usepackage{latexsym}

\usepackage[T1]{fontenc}

\usepackage[utf8]{inputenc}

\usepackage{microtype}

\usepackage{inconsolata}

\usepackage{graphicx}

\usepackage{amsmath,amssymb,booktabs,array,xcolor}
\hypersetup{pdfauthor={},pdftitle={LoCA: Local Credit Assignment for Reinforcement Learning of Autoregressive Flow-Matching TTS},pdfsubject={Local credit assignment for autoregressive flow-matching TTS}}
\newcommand{\papertitlefont}{\fontsize{15pt}{18pt}\selectfont}
\title{\papertitlefont LoCA: Local Credit Assignment for Reinforcement Learning\\of Autoregressive Flow-Matching TTS}
\input{authors.tex}
\begin{document}
\raggedbottom
\maketitle

\begin{abstract}
Reinforcement learning can improve text-to-speech synthesis through feedback on generated speech. However, sentence-level rewards do not identify local errors, while signed regression updates can lead to excessive fitting or repulsion of sampled outputs. We propose LoCA (Local Credit Assignment) for autoregressive flow-matching text-to-speech. LoCA uses ASR-based text alignment to locate content errors in generated speech and redistribute sentence-level advantages across acoustic frames. To improve training stability, we introduce reference-relative positive and negative thresholds that limit continued fitting and repulsion. Experiments on VoxCPM2 show that LoCA reduces sample-weighted content error on a multilingual CV3 evaluation subset by 20.5\% and 14.3\% relative to the pretrained backbone and GROW, respectively, while improving speaker similarity. LoCA also improves content accuracy on the Seed-TTS-Eval hard subset. Ablation results support the contributions of local credit and threshold-based regression control.
\end{abstract}

\section{Introduction}
Speech synthesis still produces mispronunciations, omissions and repetitions, motivating post-training with generated-speech feedback \citep{seedtts,glmtts}. Continuous-latent models such as DiTAR \citep{ditar} and the VoxCPM family \citep{voxcpm,voxcpm2} combine autoregressive context with diffusion or flow-based acoustics; their regression objectives provide an interface for reward-driven optimization.

This setting raises two questions. First, where should a sentence reward act? A single mispronounced word or silent interval may account for an utterance's low score, while broadcasting one coefficient treats that span and its surrounding speech identically. Second, when should a regression update stop? Positive advantages encourage fitting sampled trajectories, whereas negative advantages encourage increasing their regression error. A reference anchor opposes drift but does not specify separate stopping conditions for fitting and repulsion.

We propose LoCA (Local Credit Assignment) to address both questions. LoCA centers available frame labels within each utterance and adds them to its sentence advantage. This redistributes learning coefficients while preserving their utterance mean. For the second, we introduce reference-relative positive and negative thresholds. An improvement budget limits continued fitting, an upper error threshold stops continued repulsion, and the reference penalty remains active. Together, these components enable LoCA to target higher synthesis accuracy and more stable training without changing inference.

Evaluation should also distinguish average accuracy from severe failures. Mean CER/WER alone does not reveal how often synthesis produces silence, major content errors or truncated outputs. We define an automatic severe failure rate (SFR) using explicit content-error, silence and generation-failure conditions. An utterance satisfying any condition is counted once; SFR is the proportion of flagged utterances in the full evaluation set. We use SFR alongside CER/WER as a core metric for comparing synthesis performance across methods and training stages.

Our contributions are local credit for targeted content learning and two-sided regression control for quality improvement and reduced drift. Baseline comparisons and component ablations on VoxCPM2 examine both components using hard-set trajectories, selected multilingual CV3 splits and standard Seed evaluation.

\section{Related Work}
Speech post-training includes multi-reward optimization in Seed-TTS \citep{seedtts}, GRPO in GLM-TTS \citep{glmtts}, differentiable reward optimization in DiffRO \citep{diffro}, and token-level preference learning in TKTO \citep{tkto}. These methods motivate targeted learning from synthesis feedback; our setting uses continuous acoustic regression rather than discrete-token policy updates.

GROW \citep{grow} directly weights flow-matching regression with group-relative sentence advantages and anchors velocities to a frozen reference. FlowAWR \citep{flowawr}, evaluated in image generation, derives an advantage-rectified velocity target from a KL-constrained reward objective; AdvantageFlow \citep{advantageflow} regularizes advantage-weighted regression through the rollout policy. We build on the direct-regression approach, introducing within-utterance credit and explicit positive/negative stopping thresholds instead of changing the velocity target. GROW's existing reference anchor is retained.

\section{LoCA: Local Credit with Two-Sided Control}
\subsection{From sentence feedback to signed regression}
For each text and reference-voice condition, generate $G$ candidates, score their audio, and retain their latents for replay. Let $r_i^c=\exp(-2.5e_i)$ be the content reward and $r_i^s$ speaker similarity. Using group means and sample standard deviations, define
\begin{equation}\label{eq:adv}
A_i=\frac{r_i^c-\mu_c}{\sigma_c+10^{-8}}
 +0.5\frac{r_i^s-\mu_s}{\sigma_s+10^{-8}}.
\end{equation}
Separate normalization controls the reward scales; constant components contribute zero.

For generated latents $z$ and Gaussian noise $\epsilon$, the data-at-zero flow path \citep{flowmatching} is
\begin{equation}\label{eq:path}
x_\tau=(1-\tau)z+\tau\epsilon,\qquad y=\epsilon-z.
\end{equation}
The policy velocity $v_{it}$ and frozen pretrained reference $v^{\rm ref}_{it}$ receive the same generated history, noisy input and flow time. For channel dimension $d$, let
\begin{equation}\label{eq:errors}
\begin{aligned}
E_{it}&=\|v_{it}-y_{it}\|^2/d,\\
E^{\rm ref}_{it}&=\|v^{\rm ref}_{it}-y_{it}\|^2/d,\\
D_{it}&=\|v_{it}-v^{\rm ref}_{it}\|^2/d.
\end{aligned}
\end{equation}
Following GROW \citep{grow}, the sentence-weighted starting point is
\begin{equation}\label{eq:signed}
\mathcal L_{\rm GROW}=\frac1N\sum_{i,t}(A_iE_{it}+\beta D_{it}),
\end{equation}
where $N$ counts generated frames. Its reward term has derivative $A_i$ with respect to $E_{it}$ before averaging: the sign selects fitting or repulsion, but every frame of an utterance shares that sign and coefficient.

\subsection{Constructing and applying local credit}
\begin{figure*}[t]
\centering
\includegraphics[width=\textwidth]{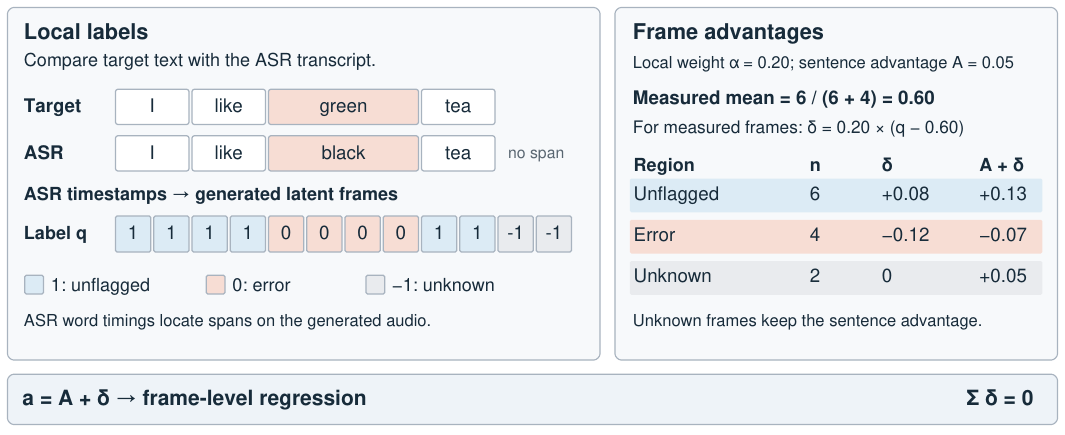}
\caption{Local-credit construction in an illustrative 12-frame example. Six unflagged and four error frames give mean label 0.6; two unknown tail frames are excluded. With local weight 0.2 and sentence advantage 0.05, error frames receive final coefficient -0.07, unflagged frames +0.13, and unknown frames retain +0.05. The boxes represent generated latent frames, not discrete speech tokens.}
\label{fig:local_credit}
\end{figure*}
\paragraph{From text errors to frame labels.}
We transcribe each generated utterance with Qwen3-ASR \citep{qwen3asr} and align the recognized words to its waveform with the forced aligner. Edit alignment against the target text identifies substitutions and insertions; their timestamps locate error spans. Deletions use neighboring-word boundaries because omitted words have no emitted span. These intervals are projected onto the exact generated latent-frame timeline, rather than treated as acoustic tokens.

\paragraph{Availability-aware feedback.}
Each frame receives $q_{it}\in\{-1,0,1\}$: unknown, detected error, or unflagged. Unknown includes unreliable timing and the unmeasured tail; it is not a negative reward. Confirmed anomalous silence is labeled 0, while eligible exact-phonetic substitutions become unknown. Define $M_{it}=\mathbf{1}[q_{it}\geq0]$. We center only measured labels within each complete utterance:
\begin{equation}\label{eq:credit}
\begin{aligned}
\bar q_i&=\frac{\sum_tM_{it}q_{it}}{\max(1,\sum_tM_{it})},\\
\delta_{it}&=\alpha M_{it}(q_{it}-\bar q_i),\qquad
a_{it}=A_i+\delta_{it}.
\end{aligned}
\end{equation}
Error and unflagged frames receive relative negative and positive increments; unknown frames receive zero. The increments sum to zero, preserving the sentence coefficient's mean. Uniform or wholly unknown labels yield zero increments. Centering is per utterance, not across candidates or microbatches.

\paragraph{Integration with regression.}
We broadcast $A_i$ to the generated frames and replace it by $a_{it}$ in the regression objective below. Labels and coefficients are detached; prefix, padding and stop positions are excluded. No separate local reward loss is added. The final coefficient need not change sign, but can do so when local evidence outweighs the sentence advantage (Figure~\ref{fig:local_credit}). Appendix~\ref{app:labels} specifies timing, deletion and label-priority rules.

\subsection{Two-sided thresholds for controlled updates}
We next control how long either sign of regression remains active. With positive budget $m_+$, negative margin $m_-$ and relative fraction $0<\rho\leq1$, define
\begin{equation}\label{eq:budget}
\begin{aligned}
b^+_{it}&=\min(m_+,\rho E^{\rm ref}_{it}),\\
\ell_{it}&=E^{\rm ref}_{it}-b^+_{it},\qquad
u_{it}=E^{\rm ref}_{it}+m_-.
\end{aligned}
\end{equation}
The relative budget keeps the positive floor nonnegative even when reference error is small. Using $[x]_+=\max(x,0)$, the combined objective is
\begin{equation}\label{eq:loss}
\begin{aligned}
\mathcal L=\frac1N\sum_{i,t}\big(&[a_{it}]_+[E_{it}-\ell_{it}]_+\\
&+[-a_{it}]_+[u_{it}-E_{it}]_++\beta D_{it}\big).
\end{aligned}
\end{equation}
Feedback, targets and reference predictions are detached. Each active branch retains the signed-regression derivative: positive fitting stops at the improvement budget, and negative repulsion stops at the upper threshold. The reference penalty continues to anchor both branches. This removes the incentive for unbounded negative regression and limits repeated positive fitting. Appendix~\ref{app:analysis} derives the gradients and fixed-replay bounds; these properties concern the regression objective rather than a global bound on parameter drift.

\section{Experiments}
\begin{table*}[t]
\centering\fontsize{10pt}{12pt}\selectfont\setlength{\tabcolsep}{3pt}
\input{tables/main.tex}
\caption{Main model comparison. Every metric in a row uses the same selected model. Cases are severe proxy counts and SIM is cosine similarity scaled by 100. Lower Error/Cases and higher SIM are preferred. CV3 regular covers the 3,500 ordinary-language items; the full selection also includes 124 hard items. CV3 hard weights 60 Chinese and 64 English items.}
\label{tab:main}
\end{table*}
\begin{table*}[t]
\centering\fontsize{10pt}{12pt}\selectfont\setlength{\tabcolsep}{4pt}
\input{tables/ablation.tex}
\caption{LoCA component ablations. Yes/No indicates component presence. Metrics use the full 3,624-item CV3 selection. The variant w/o thresholds removes the positive and negative regression thresholds.}
\label{tab:ablation}
\end{table*}
\subsection{Backbone, data and evaluation}
All systems use the publicly released 2B-parameter VoxCPM2 checkpoint (\texttt{openbmb/VoxCPM2}; \citealp{voxcpm2repo}). The base is evaluated without RL; GROW, LoCA and the ablations start from this same checkpoint.

RL samples a 50,000-example Emilia \citep{emilia} pool covering Chinese, English, German, French, Japanese and Korean. Hyperparameters are in Appendix~\ref{app:protocol}. Evaluation uses seven CV3 \citep{cv3eval} language splits (Chinese, English, German, Spanish, French, Italian and Russian; 500 each), plus hard Chinese/English (60/64), totaling 3,624. This is a fixed subset of CV3. Seed-TTS-Eval \citep{seedttseval} includes Chinese hard (400) and standard Chinese/English (2,020/1,088).

We select representative models by jointly considering content accuracy, severe failures and speaker similarity. Complete training trajectories and supplementary results are provided in Appendix~\ref{app:trajectory}.

\subsection{Automatic severe failure rate}
For item $j$, let $e_j$ be utterance CER/WER; $S_j$, $C_j$ and $X_j$ indicate all-silent output, a generation-cap hit and a generation error. Let $T_j$ and $I_j$ be tail and longest internal silence in seconds:
\begin{equation}\label{eq:sfr}
\begin{aligned}
F_j&=\mathbf{1}[e_j>0.5\lor S_j\lor C_j\lor X_j\\
&\qquad\lor T_j\geq1.2\lor I_j\geq1.5],\\
\mathrm{SFR}&=\frac{100}{M}\sum_{j=1}^{M}F_j.
\end{aligned}
\end{equation}
The denominator $M$ is the full evaluation set, including generation errors. Each flagged item counts once; tables and plots report this union count as severe proxy cases. Detection details are in Appendix~\ref{app:failures}.

\subsection{Main results and ablations}
LoCA achieves CV3 error 4.758\%, versus 5.986\% for the base and 5.553\% for GROW (Table~\ref{tab:main}). Severe proxy cases decrease from 61/60 to 51, and SIM rises from 67.168/67.562 to 69.137. On ordinary CV3, LoCA has error 4.673\%, versus 5.887\%/5.461\%. It also lowers Seed-hard CER to 6.884\% with 5 severe cases. CV3-hard error is 7.163\%, with 6 cases: lower error than both baselines, with the base's case count and one fewer case than GROW.

Table~\ref{tab:ablation} evaluates component removals. Without local credit, CV3 error is 5.508\%, with 54 cases and SIM 67.525. Without both thresholds, it is 5.488\%, with 44 cases and SIM 67.722. Full LoCA has better content accuracy and speaker similarity than both variants; the no-threshold variant has fewer severe cases. The trajectories retain these metric-dependent trade-offs.

\subsection{Russian generalization}
Russian is excluded from the six-language RL pool. On its 500-item CV3 split, LoCA's selected checkpoint reduces WER from the base's 6.219\% to 4.723\% (24.1\% relative) and raises SIM from 67.296 to 68.541; severe proxy cases are 6 versus the base's 5. WER remains below the base at every evaluated LoCA checkpoint. This supports transfer beyond the RL training languages; it does not imply Russian was absent from backbone pretraining. Language profiles and similarity trajectories are in Appendix~\ref{app:cv3}.

\section{Conclusion}
LoCA combines local credit with two-sided regression control for continuous TTS. The selected model improves CV3 and Seed-hard content accuracy over the baselines, with higher CV3 speaker similarity. Russian results show content and similarity gains outside the RL training pool. Full trajectories retain the cross-checkpoint and hard-set trade-offs.

\section*{Limitations}
Experiments use one training seed and one backbone. Results describe the evaluated language and task subsets.

Local labels inherit recognition and timing errors. Phonetic matching is a text-side proxy, while the severe-failure metric is an automatic diagnostic. Fixed-replay threshold properties do not guarantee optimizer convergence or absence of parameter drift.

\section*{Ethical Considerations}
Voice cloning can enable impersonation. Audio/model release requires appropriate dataset permissions and speaker rights; this study does not establish redistribution rights for every source recording. Automatic error measures do not guarantee fairness across accents or dialects.

\bibliography{references}
\appendix
\renewcommand{\dbltopfraction}{0.82}
\renewcommand{\textfraction}{0.12}
\makeatletter
\setlength{\@fptop}{0pt}
\setlength{\@dblfptop}{0pt}
\setlength{\@fpsep}{12pt}
\setlength{\@dblfpsep}{12pt}
\makeatother

\section{Training and Evaluation Protocol}\label{app:protocol}
\paragraph{Training.}
All trained configurations start from the BF16 release of the 2B-parameter VoxCPM2 model \citep{voxcpm2,voxcpm2repo}; GROW denotes its advantage-weighted objective applied to this backbone. The 50,000-example Emilia pool spans Chinese, English, German, French, Japanese and Korean. Pool size is distinct from training duration, which is specified by steps. We use group size 8, replay microbatches of two, and two-condition gradient accumulation. AdamW \citep{adamw} uses learning rate $2\times10^{-6}$, 50-step warmup, zero weight decay, gradient clipping at 0.5 and seed 42. LoCA uses $\alpha=.2$, $m_+=m_-=.1$, $\rho=.5$ and $\beta=.025$.

Training samples use ten flow steps, classifier-free guidance \citep{cfg} scale 2 and a 20-second duration ceiling. Generated latents are retained for replay. The audio autoencoder, patch encoder and stop head are frozen; no supervised fine-tuning or stop loss is added. Losses are averaged over generated frames. Content rewards use Qwen3-ASR \citep{qwen3asr}; speaker rewards use similarity to the reference voice.

\paragraph{Evaluation.}
All systems use reference-audio conditioning, ten flow steps, guidance scale 2 and fixed per-example seeds, without retries or best-of-$N$ selection. CV3 includes only the seven ordinary splits and two hard splits listed in the main text; it excludes Japanese/Korean and other CV3 tasks. Russian, Spanish and Italian are outside the RL training language pool.

Chinese uses a Paraformer-family recognizer \citep{paraformer}; other evaluated languages use Whisper large-v3 \citep{whisper}. Content error is the untruncated mean of utterance CER/WER. CV3 aggregates these rates by sample count. SIM is speaker-embedding cosine similarity times 100, using ERes2Net for CV3 \citep{eres2net} and the benchmark's fine-tuned WavLM verifier for Seed \citep{wavlm,seedttseval}. DNSMOS-P808 is an auxiliary automatic quality estimate \citep{dnsmos}, not human MOS.

\section{Local Labels and Objective Properties}\label{app:labels}

The alignment procedure compares normalized reference and ASR transcripts, then uses timestamps aligned to the ASR hypothesis. For waveform duration $D$ and $L$ generated latent frames, an error interval from $s$ to $e$ maps to frame indices from $\lfloor sL/D\rfloor$ to $\lceil eL/D\rceil$ (end exclusive), clipped to the generated range. The mask contains exactly $L$ generated-frame entries.

Measured regions start at label 1; substitutions and insertions overwrite their aligned spans with 0. Interior deletions mark the boundary between adjacent recognized tokens, using an 80-ms window when no gap exists. Beginning/end deletions use an 80-ms window clipped to the neighboring word. These are localization heuristics, not timestamps of an emitted missing word. Frames beyond the final reliable timestamp are unknown ($-1$). Invalid alignment, near-silent audio or generation-cap hits disable ASR labels; independently confirmed silence can still supply local evidence.

Silence labels require an anomalous-silence detection supported by waveform energy. A 0.3-second grace interval avoids penalizing uncertain boundaries. Only frames fully inside the retained silence interval are marked 0. These acoustic labels override ASR labels, including phonetic-neutral regions. The recognition/timestamp helpers build on public GLM-TTS code \citep{glmttscode}; the within-utterance centered construction is the method studied here.

Phonetic neutralization applies only to equal-length substitution chunks with nonempty, exactly matching pronunciation sequences and valid timestamps. Eligible error labels become unavailable, preserving overlapping insertion/deletion penalties and acoustic overlays. Uniform all-error labels have no within-utterance contrast and hence zero local increment; sentence feedback may still update those candidates.

\subsection{Fixed-replay derivations}\label{app:analysis}
\paragraph{Mean-preserving local credit.}
For an utterance with $n_i=\sum_t M_{it}>0$, substituting Eq.~\ref{eq:credit} gives
\begin{equation}\label{eq:centerproof}
\sum_t\delta_{it}=\alpha\left(\sum_tM_{it}q_{it}-n_i\bar q_i\right)=0.
\end{equation}
When $n_i=0$, every increment is zero by the mask. If the fraction of unflagged measured frames is $p$, error and unflagged frames receive increments $-\alpha p$ and $\alpha(1-p)$, respectively. This preserves the mean coefficient before thresholding; different residuals and network Jacobians prevent interpreting it as gradient conservation.

\paragraph{Why an anchor alone need not bound signed regression.}
Fix one replay frame, its coefficient $a$, target $y$ and reference $v^{\rm ref}$. Expanding the unaveraged signed loss yields
\begin{equation}\label{eq:quadratic}
\begin{aligned}
L_{\rm signed}(v)&=a\|v-y\|^2/d+\beta\|v-v^{\rm ref}\|^2/d\\
&=\frac{a+\beta}{d}\|v\|^2\\
&\quad-\frac{2}{d}v^\top(ay+\beta v^{\rm ref})+C,
\end{aligned}
\end{equation}
where $C$ is independent of $v$. If $a+\beta<0$, the quadratic tends to negative infinity as the output norm grows. This is a sufficient condition for a problematic frame objective, not a claim that every GROW update diverges.

\paragraph{Thresholded derivatives.}
Let $h(E)=[a]_+[E-\ell]_++[-a]_+[u-E]_+$ be the unaveraged RL term. Away from the hinge boundaries,
\begin{equation}\label{eq:gategrad}
\frac{\partial h}{\partial E}=
\begin{cases}
a\,\mathbf{1}[E>\ell],&a>0,\\
a\,\mathbf{1}[E<u],&a<0,\\
0,&a=0.
\end{cases}
\end{equation}
Hence each active branch matches signed regression, and its RL gradient vanishes after saturation. Because $\nabla_v E=2(v-y)/d$, the active negative branch satisfies
\begin{equation}\label{eq:negativebound}
\|\nabla_v h\|=2|a|\sqrt{E/d}
 \leq 2|a|\sqrt{(E^{\rm ref}+m_-)/d}.
\end{equation}
All terms of the thresholded loss are nonnegative. For $\beta>0$, the per-frame total also satisfies $L(v)\geq\beta\|v-v^{\rm ref}\|^2/d$, so it grows as the output moves arbitrarily far from the fixed reference. The positive branch is not upper-bounded in error. These statements concern a fixed replay output; the hinge is nonsmooth at its boundary, and shared parameters can move a saturated frame again. They establish control of the regression terms, not optimizer convergence or globally drift-free training.

\section{Checkpoint Trajectories and Compact Summaries}\label{app:trajectory}
\begin{figure*}[t]
\centering
\includegraphics[width=\textwidth]{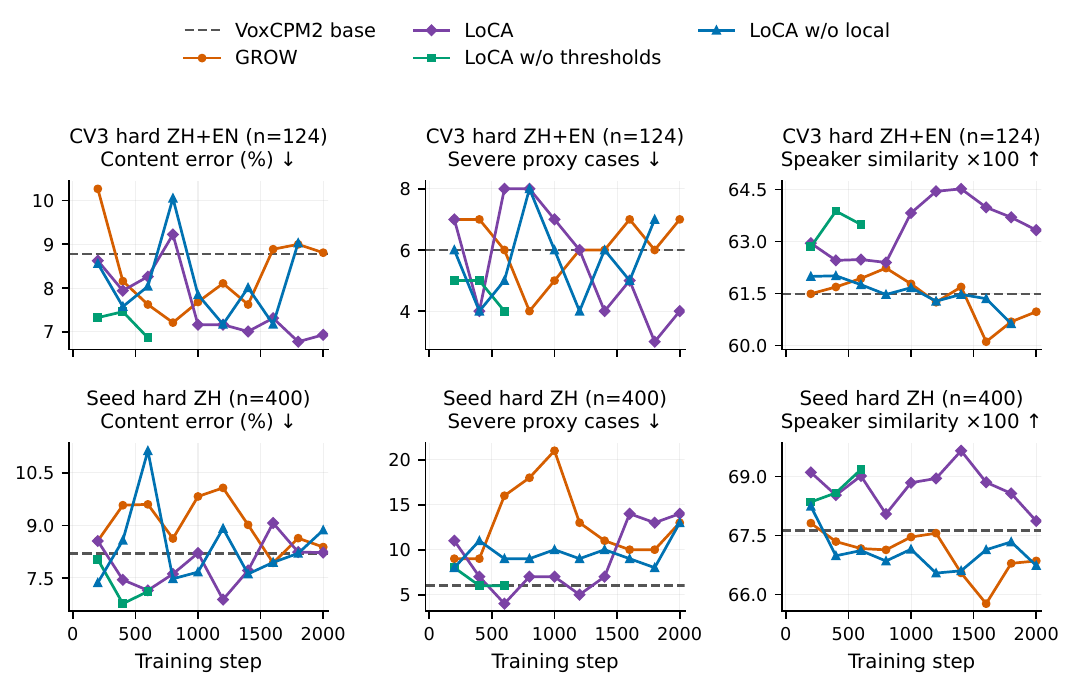}
\caption{Hard-set trajectories for the baselines and LoCA ablations. Columns show content error, severe proxy cases and speaker similarity. Chinese/English CV3 hard means use weights 60/124 and 64/124; case counts are summed. The dashed line is the base. Lines connect evaluated points without smoothing or extrapolation.}
\label{fig:hard_trajectories}
\end{figure*}

Figure~\ref{fig:hard_trajectories} retains the hard-set training trajectories through step 2000. CV3 hard combines the 60 Chinese and 64 English items: content error and SIM use sample weights, whereas severe case counts are summed. Each group ends at its last evaluated point; in particular, the no-local CV3 series ends at 1800 while its Seed-hard series reaches 2000.

Each model's main-table metrics, language scores and radar profile use the same selected checkpoint. The trajectories show how all three metrics vary during training.

Standard Seed has a separate selection rule: within each method and language, choose the lowest observed CER/WER and report the other metrics at that same step (Table~\ref{tab:standard_seed}). English and Chinese steps can differ and are explicitly listed; no row combines metrics from different checkpoints.

\begin{table*}[t]
\centering\fontsize{9.5pt}{11pt}\selectfont\setlength{\tabcolsep}{3pt}
\input{tables/seed_standard.tex}
\caption{Best observed standard Seed results per method and language. Each row is selected by its own content error, with step shown. Error is CER for ZH and WER for EN; Cases is the severe proxy count and SIM is scaled by 100.}
\label{tab:standard_seed}
\end{table*}

\section{Severe Failure Categories and Detection}\label{app:failures}
Table~\ref{tab:failures} shows Seed-hard categories at the reported checkpoints. High error means utterance CER above 50\%; tail and internal silence thresholds are 1.2 and 1.5 seconds. Categories overlap, so their counts do not sum to the union. Leading silence and recognition failures alone do not trigger SFR.

\begin{table*}[t]
\centering\fontsize{9.5pt}{11pt}\selectfont\setlength{\tabcolsep}{3pt}
\input{tables/failures.tex}
\caption{Seed-hard severe-failure categories at each method's reported checkpoint, $n=400$. Union counts each flagged utterance once.}
\label{tab:failures}
\end{table*}

Silence detection uses mono audio, a 50-ms RMS window and a 20-ms hop. The activity threshold is 15\% of the waveform's 95th-percentile frame RMS $q_{95}$, clipped to $[0.0003,0.003]$. No active frames means all-silent output; tail and internal durations are measured from active-frame positions. A cap hit denotes maximum-length stopping. These flags are reproducible diagnostics: a cap hit need not imply omitted content, and ASR differences can include phonetic or formatting ambiguities.

\section{Multilingual Generalization and Similarity}\label{app:cv3}
\begin{figure*}[t]
\centering
\includegraphics[width=\textwidth]{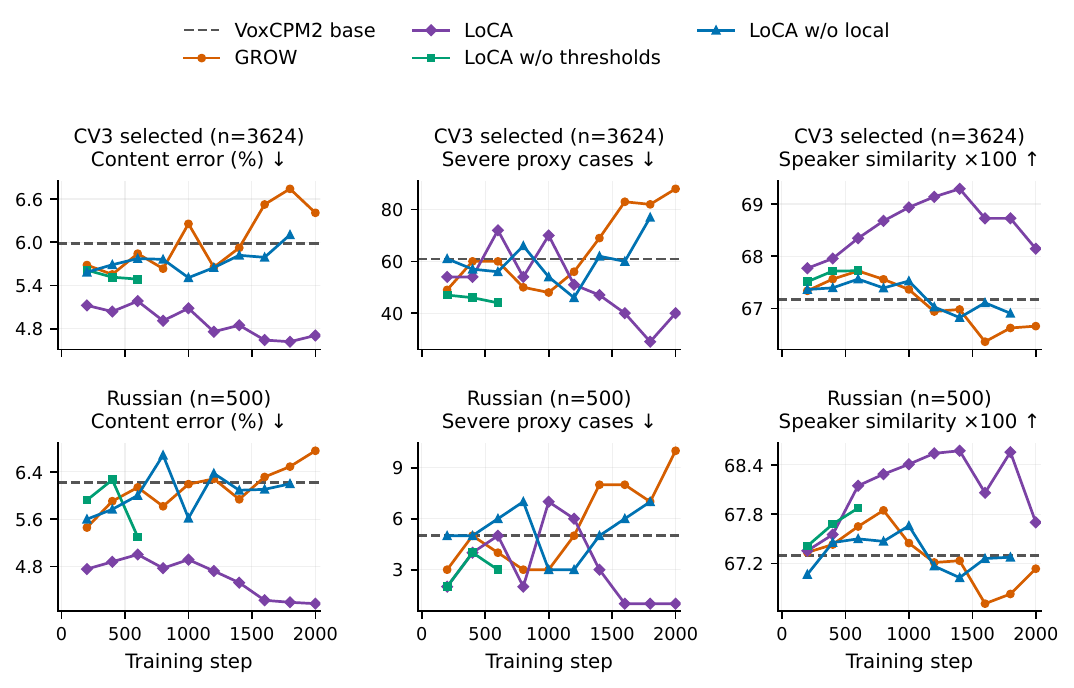}
\caption{Full-subset CV3 and Russian trajectories, including speaker similarity. Russian is outside the six-language RL training pool. All plotted points are reported evaluations; the base is a horizontal reference.}
\label{fig:cv3_russian}
\end{figure*}

Figure~\ref{fig:cv3_russian} reports all three metrics for the full CV3 selection and Russian. Russian was excluded from RL training, while the pretrained backbone can already synthesize it. At the reported checkpoints, Russian WER is 6.219\% for the base, 5.904\% for GROW and 4.723\% for LoCA. The improvements therefore reflect post-training transfer rather than first acquisition of the language.

\begin{table*}[t]
\centering\fontsize{9.5pt}{11pt}\selectfont\setlength{\tabcolsep}{3pt}
\input{tables/quality.tex}
\caption{Auxiliary quality metrics at the selected checkpoints and representative training steps for GROW and LoCA. SIM and P808 are higher-is-better; SIM backends differ between CV3 and Seed.}
\label{tab:quality}
\end{table*}

\begin{figure*}[t]
\centering
\includegraphics[width=.78\textwidth]{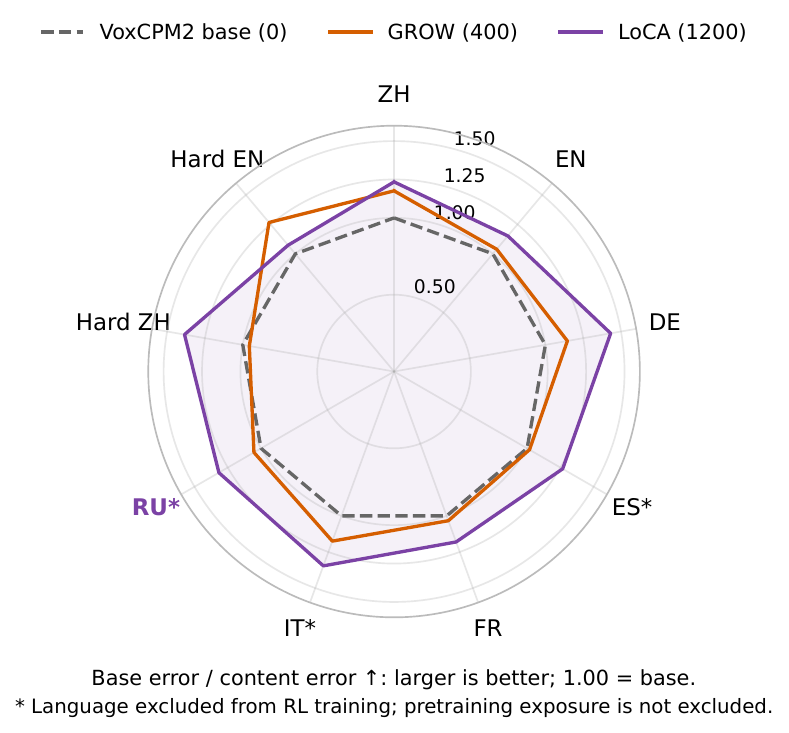}
\caption{Inverse error ratio (base error / method error), at each system's reported checkpoint. Larger is better; 1 is the base. Asterisks mark Spanish, Italian and Russian, excluded from the RL training pool. Hard Chinese/English are separate tasks, not extra languages.}
\label{fig:language_radar}
\end{figure*}

\newpage
The radar chart divides each split's base error by the method's corresponding error (Figure~\ref{fig:language_radar}). This avoids directly mixing language-dependent CER/WER scales; a radius above one indicates improvement. The chart is a profile overview, not an aggregate score or a claim based on polygon area. Exact errors and checkpoint steps are retained in Table~\ref{tab:languages}.

\begin{table*}[t]
\centering\fontsize{9.5pt}{11pt}\selectfont\setlength{\tabcolsep}{2.5pt}
\input{tables/languages.tex}
\caption{Exact CV3 language errors at each method's selected checkpoint. Chinese uses CER; other splits use WER. All values are percentages; h-zh/h-en are the hard subsets.}
\label{tab:languages}
\end{table*}

\end{document}

%% file: authors.tex
\author{{\bfseries Jie Liu$^{*}$ \quad Junyu Zhou \quad Junhao Zhong \quad WeiSen Tang} \\ {\bfseries Rukai Cai \quad Haoyue Zhan \quad Ruiqi Xing \quad Ling Long} \\ {\normalfont Lingguang Zhaxian Technology, Shanghai, China} \\ {\normalfont\small $^{*}$Corresponding author: \href{mailto:liujie1@52tt.com}{\texttt{liujie1@52tt.com}}}}
\hypersetup{pdfauthor={Jie Liu, Junyu Zhou, Junhao Zhong, WeiSen Tang, Rukai Cai, Haoyue Zhan, Ruiqi Xing, Ling Long}}

%% file: tables/main.tex
\begin{tabular}{lrrrrrrrr}
\toprule
\multicolumn{1}{c}{} & \multicolumn{3}{c}{CV3 selected (3,624)} & CV3 regular & \multicolumn{2}{c}{Seed hard} & \multicolumn{2}{c}{CV3 hard} \\
\cmidrule(lr){2-4}\cmidrule(lr){5-5}\cmidrule(lr){6-7}\cmidrule(lr){8-9}
Method & Error $\downarrow$ & Cases $\downarrow$ & SIM $\uparrow$ & Error $\downarrow$ & Error $\downarrow$ & Cases $\downarrow$ & Error $\downarrow$ & Cases $\downarrow$ \\
\midrule
VoxCPM2 base & 5.986 & 61 & 67.168 & 5.887 & 8.193 & 6 & 8.772 & 6 \\
GROW & 5.553 & 60 & 67.562 & 5.461 & 9.575 & 9 & 8.157 & 7 \\
LoCA & 4.758 & 51 & 69.137 & 4.673 & 6.884 & 5 & 7.163 & 6 \\
\bottomrule
\end{tabular}

%% file: tables/ablation.tex
\begin{tabular}{lccrrr}
\toprule
LoCA variant & Local & Thresholds & Error $\downarrow$ & Cases $\downarrow$ & SIM $\uparrow$ \\
\midrule
LoCA (full) & Yes & Yes & 4.758 & 51 & 69.137 \\
w/o local credit & No & Yes & 5.508 & 54 & 67.525 \\
w/o thresholds & Yes & No & 5.488 & 44 & 67.722 \\
\bottomrule
\end{tabular}

%% file: tables/seed_standard.tex
\begin{tabular}{llrrrrr}
\toprule
Method & Split & Step & n & Error $\downarrow$ & Cases $\downarrow$ & SIM $\uparrow$ \\
\midrule
VoxCPM2 base & EN & 0 & 1088 & 1.710 & 1 & 66.789 \\
VoxCPM2 base & ZH & 0 & 2020 & 1.006 & 1 & 72.676 \\
GROW & EN & 800 & 1088 & 1.621 & 2 & 67.588 \\
GROW & ZH & 800 & 2020 & 0.911 & 3 & 73.361 \\
LoCA w/o thresholds & EN & 400 & 1088 & 1.544 & 0 & 67.336 \\
LoCA w/o thresholds & ZH & 600 & 2020 & 0.914 & 0 & 74.140 \\
LoCA & EN & 800 & 1088 & 1.530 & 0 & 69.362 \\
LoCA & ZH & 400 & 2020 & 0.800 & 0 & 75.101 \\
LoCA w/o local & EN & 400 & 1088 & 1.572 & 0 & 67.539 \\
LoCA w/o local & ZH & 800 & 2020 & 0.858 & 2 & 73.180 \\
\bottomrule
\end{tabular}

%% file: tables/failures.tex
\begin{tabular}{lrrrrrrr}
\toprule
Method & Step & High err. $\downarrow$ & Silent $\downarrow$ & Tail $\downarrow$ & Internal $\downarrow$ & Cap $\downarrow$ & Union $\downarrow$ \\
\midrule
VoxCPM2 base & 0 & 5 & 0 & 0 & 0 & 3 & 6 \\
GROW & 400 & 8 & 0 & 0 & 0 & 7 & 9 \\
LoCA w/o thresholds & 600 & 5 & 0 & 0 & 0 & 3 & 6 \\
LoCA & 1200 & 5 & 0 & 0 & 0 & 1 & 5 \\
LoCA w/o local & 1000 & 8 & 0 & 0 & 1 & 4 & 10 \\
\bottomrule
\end{tabular}

%% file: tables/quality.tex
\begin{tabular}{lrrrrr}
\toprule
Method & Step & CV3 SIM $\uparrow$ & CV3 P808 $\uparrow$ & Seed-h SIM $\uparrow$ & Seed-h P808 $\uparrow$ \\
\midrule
VoxCPM2 base & 0 & 67.168 & 3.570 & 67.632 & 3.691 \\
GROW & 400 & 67.562 & 3.580 & 67.346 & 3.664 \\
GROW & 800 & 67.558 & 3.556 & 67.135 & 3.626 \\
GROW & 1200 & 66.941 & 3.537 & 67.559 & 3.680 \\
GROW & 1600 & 66.363 & 3.551 & 65.767 & 3.694 \\
GROW & 2000 & 66.662 & 3.581 & 66.854 & 3.640 \\
LoCA w/o thresholds & 600 & 67.722 & 3.605 & 69.179 & 3.753 \\
LoCA & 800 & 68.676 & 3.601 & 68.046 & 3.666 \\
LoCA & 1200 & 69.137 & 3.636 & 68.951 & 3.641 \\
LoCA & 1600 & 68.724 & 3.619 & 68.853 & 3.649 \\
LoCA & 2000 & 68.144 & 3.604 & 67.868 & 3.698 \\
LoCA w/o local & 1000 & 67.525 & 3.564 & 67.151 & 3.648 \\
\bottomrule
\end{tabular}

%% file: tables/languages.tex
\begin{tabular}{lrrrrrrrrrr}
\toprule
Method & Step & zh $\downarrow$ & en $\downarrow$ & de $\downarrow$ & es $\downarrow$ & fr $\downarrow$ & it $\downarrow$ & ru $\downarrow$ & h-zh $\downarrow$ & h-en $\downarrow$ \\
\midrule
VoxCPM2 base & 0 & 3.996 & 5.193 & 5.551 & 4.571 & 10.657 & 5.023 & 6.219 & 9.952 & 7.665 \\
GROW & 400 & 3.398 & 4.997 & 4.846 & 4.491 & 10.314 & 4.275 & 5.904 & 10.402 & 6.053 \\
LoCA w/o thresholds & 600 & 3.396 & 5.017 & 4.500 & 5.306 & 10.638 & 3.917 & 5.300 & 8.501 & 5.333 \\
LoCA & 1200 & 3.238 & 4.507 & 3.878 & 3.608 & 9.027 & 3.731 & 4.723 & 7.186 & 7.143 \\
LoCA w/o local & 1000 & 3.402 & 4.806 & 4.562 & 4.284 & 10.502 & 4.803 & 5.617 & 8.120 & 7.593 \\
\bottomrule
\end{tabular}